\documentclass[conference]{IEEEtran}
\IEEEoverridecommandlockouts
\usepackage{cite}
\usepackage{amsmath,amssymb,amsfonts}
\usepackage{algorithmic}
\usepackage{graphicx}
\usepackage{textcomp}
\usepackage{xcolor}

\usepackage{graphics}
\usepackage{cite}
\usepackage{subfig}
\usepackage{stmaryrd}
\usepackage{color}
\usepackage{amsthm}
\usepackage{amsmath}
\usepackage{enumerate}
\usepackage[colorlinks=true, allcolors=blue]{hyperref}
\usepackage{booktabs}
\usepackage{url}
\usepackage{multirow}
\usepackage{bbm}
\usepackage{algorithm}
\usepackage{algorithmic}
\usepackage{bbding} 
\usepackage{threeparttable} 
\usepackage{setspace} 

\usepackage{comment}

\allowdisplaybreaks[3]

\DeclareRobustCommand*{\IEEEauthorrefmark}[1]{
    \raisebox{0pt}[0pt][0pt]{\textsuperscript{\footnotesize\ensuremath{#1}}}}

\def\BibTeX{{\rm B\kern-.05em{\sc i\kern-.025em b}\kern-.08em
    T\kern-.1667em\lower.7ex\hbox{E}\kern-.125emX}}
\begin{document}

\title{Enabling Privacy-Preserving and Publicly Auditable Federated Learning\\
}

\author{
\IEEEauthorblockN{
Huang Zeng\IEEEauthorrefmark{1},
Anjia Yang\IEEEauthorrefmark{1},
Jian Weng\IEEEauthorrefmark{1},
Min-Rong Chen\IEEEauthorrefmark{2},
Fengjun~Xiao\IEEEauthorrefmark{3,4},
Yi Liu\IEEEauthorrefmark{1}, and
Ye Yao\IEEEauthorrefmark{4}}
\IEEEauthorblockA{\IEEEauthorrefmark{1}College of Cyber Security, Jinan University, Guangzhou 510632, China}
\IEEEauthorblockA{\IEEEauthorrefmark{2}School of Computer Science, South China Normal University, Guangzhou 510631, China}
\IEEEauthorblockA{\IEEEauthorrefmark{3}Zhejiang Informatization Development Institute, Hangzhou, China}
\IEEEauthorblockA{\IEEEauthorrefmark{4}Hangzhou Dianzi University, Hangzhou, 310018, China}
\IEEEauthorblockA{\{huagzeg, anjiayang, cryptjweng\}@gmail.com, mrongchen@126.com, \{xiaofengjun, yaoye\}@hdu.edu.cn, liuyi@jnu.edu.cn  }
\IEEEauthorblockA{Corresponding Authors: Anjia Yang and Fengjun~Xiao}
}

\maketitle

\begin{abstract}
Federated learning (FL) has attracted widespread attention because it supports the joint training of models by multiple participants without moving private dataset. However, there are still many security issues in FL that deserve discussion. In this paper, we consider three major issues: 1) how to ensure that the training process can be publicly audited by any third party; 2) how to avoid the influence of malicious participants on training; 3) how to ensure that private gradients and models are not leaked to third parties. Many solutions have been proposed to address these issues, while solving the above three problems simultaneously is seldom considered. In this paper, we propose a publicly auditable and privacy-preserving federated learning scheme that is resistant to malicious participants uploading gradients with wrong directions and enables anyone to audit and verify the correctness of the training process. In particular, we design a robust aggregation algorithm capable of detecting gradients with wrong directions from malicious participants. Then, we design a random vector generation algorithm and combine it with zero sharing and blockchain technologies to make the joint training process publicly auditable, meaning anyone can verify the correctness of the training. Finally, we conduct a series of experiments, and the experimental results show that the model generated by the protocol is comparable in accuracy to the original FL approach while keeping security advantages.
\end{abstract}


\section{Introduction}

Since federated learning (FL) \cite{FL:FedSGD/McMahanM16, FL:FedAvg/McMahanM17} was proposed, it has been widely concerned and used to meet various real-world needs, such as next-word prediction \cite{FL:GKeyboard/YangA19} and medical imaging \cite{FL:MedicalAI/LiM19} because it can alleviate privacy issues and prevent data misuse in machine learning. Specifically, in FL, a central server and multiple participants (data providers) perform multiple rounds of model and gradient exchange to train the global model without sharing the training dataset.

However, there are still many security issues in federated learning that deserve our discussion. For example, it is difficult to support public auditing on the joint model training process which is easily negatively affected by malicious participants to output incorrect models. Besides, it is crucial to protect the models and gradients during joint training from being leaked to any third party, since this will harm the actual interests of central server and participants. These   
 problems will lead to participants' resistance to joining FL and users' lack of trust in the models output by FL, thus limiting the application and deployment of FL \cite{FL:Verifiable/Auditable/Blockchain/PengX22, FL:Trust/Blockchain/SinKitL22}.

In recent years, many efforts have been made to solve the above problems in FL. Some solutions use blockchain and smart contracts to perform auditing and verification operations on FL. In FLChain \cite{FL:Auditable/Blockchain/FLChain/BaoS19}, the authors build a publicly auditable federated learning where participants upload local gradients to the blockchain and download aggregated gradients calculated by smart contracts. VFChain \cite{FL:Verifiable/Auditable/Blockchain/PengX22}, which combines blockchain and federated learning, supports any node on the blockchain to audit training process based on the Chameleon hashing and signature algorithms. However, the use of transparent blockchain technology raises privacy concerns.
DeepChain \cite{FL:Auditable/Privacy/Blockchain/DeepChain/WengW21} employs homomorphic encryption, smart contracts, and trusted execution environment (TEE) to achieve privacy-preserving and publicly auditable gradient aggregation, while it relies on TEE and specific blockchain.

There are also many scholars who employ cryptographic primitives to safeguard FL. For example, by encrypting gradients and generating proofs of correct aggregation,  VerifyNet \cite{FL:Verifiable/Privacy/XuL20} and PVD-FL \cite{FL:Privacy/Verifiable/ZhaoZ22} can effectively achieve privacy protection and verifiability. Other methods \cite{FL:Privacy/Robust/XuL22,FL:Privacy/Robust/SMPC/HaoL21,FL:Privacy/KanchanC22} achieve privacy and robust aggregation through the design of specific secure computation protocols. However, they do not consider public auditability or robustness against malicious participators, such as detecting local gradients with wrong directions using cosine similarity, which is commonly used to implement robust aggregation algorithms  \cite{FL:robust/ZhangC22, FL:robust/FLtrust/CaoF21}.

Based on the above observations, we make the following contributions in this paper.
\begin{enumerate}
    \item We propose a publicly auditable and privacy-preserving federated learning scheme based on blockchain and secure multi-party computation (SMPC) technology. Even if malicious participants participate in training, the scheme can produce legitimate and high-quality global models that can be used in real-world applications. 
    \item For public auditability, we propose an $PRVG$ algorithm and adopt a robust aggregation algorithm, and then combine them with zero sharing to generate training records. This allows anyone on the blockchain to trace the training process without compromising privacy to verify the correctness of the training.
    \item Analysis shows that our scheme meets security requirements. Our experimental evaluation of the scheme shows that our scheme can produce models with high accuracy comparable to the original federated learning approach.
\end{enumerate}

The structure of this article is as follows. The system model and cryptographic primitives are presented in section \ref{section:pre}. The proposed federated model training scheme and security analysis are described in section \ref{section:our_scheme}. We give the evaluation in section \ref{section:evaluation} and we conclude this paper in section \ref{section:conclusion}.

\section{Preliminaries}\label{section:pre}

In this section, we present the system model and briefly introduce the fundamental security primitives that contribute to building the the proposed scheme.

\subsection{System Model}

\begin{figure}[htbp]
\centering
\includegraphics[width=0.75\linewidth]{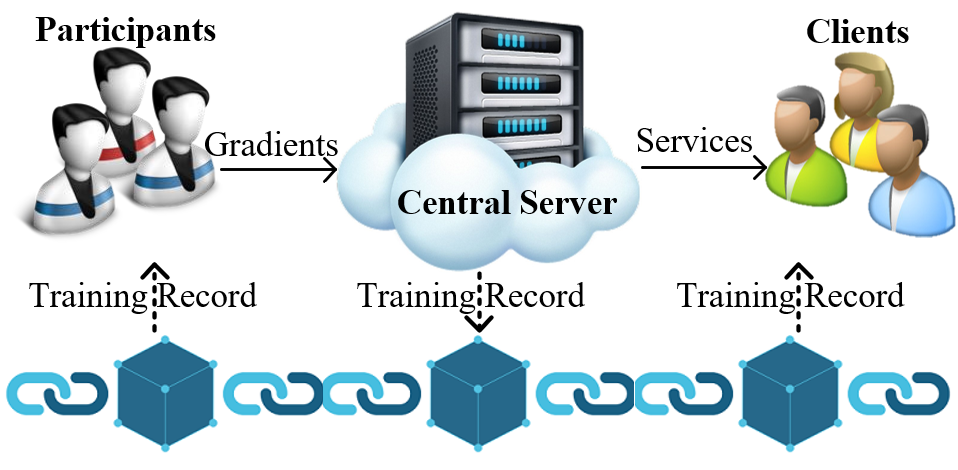}
\caption{System Model.}
\label{figure:systemmodel}
\end{figure}
\vspace{0cm}

As shown in Figure \ref{figure:systemmodel}, the central server $CS$ formulates model training goals and performs FL with participants. In the $r$-th round of training, $CS$ first releases the global model $\mathbf{w}_{r-1}$ to participants. Each participant $i$ executes the stochastic gradient descent algorithm to obtain local gradient $\mathbf{g}^i_r$ and sends it to $CS$. $CS$ then inputs all received local gradients into the aggregation algorithm to obtain an aggregated gradient $\mathbf{ag}_r$ and updates the global model.

After training is completed, $CS$ will deploy the model in inference services, upload the training record to the blockchain. Clients receiving services can audit the training process through the training record on the blockchain.

\textbf{\textit{Threat Model:}} $CS$ is honest and reliable for the participants, but not for the clients. Participants may be malicious and they want to obtain training data from other parties. In particular, in order to destroy the training process, malicious participants will modify the label of training data or falsify to provide a bad local gradient with wrong directions. The client will want to get the global model so that he does not have to pay for the services provided by $CS$. We adopt the standard security model in many blockchain systems, considering that at least $2/3$ of the blockchain nodes are honest.

\textbf{\textit{Design goals:}} The training data of all parties and the data label set ($L_S$ of $CS$ and $L_i$ of participant $i$) should not be disclosed to anyone. Secondly, the global model as the private property of $CS$ can not be disclosed. At the same time, the scheme needs to ensure the robustness of the training process, i.e. bad gradients uploaded by participants do not affect the accuracy of the global model. Finally, the training process should also be able to be audited by any third party, such as clients, to verify the correctness of the training process.

\subsection{Security Primitives}

Following are the security primitives required for the scheme.

\subsubsection{Zero Sharing} As a type of additive secret sharing, zero sharing is widely used in SMPC. The idea of this primitive is that one party divides $0$ into multiple secret shares and sends them to other parties ($z_i$ for party $i$), the sum of these shares equals $0$. It is worth noting that zero sharing, like additive secret sharing, has additive homomorphism.

Assuming that $s^i$ is the share of secret value $s$ held by party $i\in P$, and that $s$ can be recovered by adding $|P|$ shares. Let each party calculates $\Tilde{s}^i=s^i+z^i$, then according to the homomorphism of the additive secret shares, we can get $\Tilde{s}=\sum_{i\in P}\Tilde{s}^i=\sum_{i\in P}(s^i+z^i)=s+0=s$.

\subsubsection{Elliptic Curve Digital Signature (ECDSA) Algorithm} Digital signatures have the characteristics of non-repudiation, non-tampering and non-forgery and supports public verification. The ECDSA algorithm includes three algorithms:
\begin{itemize}
    \item $(sk,pk)\gets KeyGen(1^k)$ takes security parameter $1^k$ and elliptic curve group $G$ as input and outputs a pair of private keys $(sk,pk)$.
    \item $\sigma\gets Sign(sk,m)$ takes secret key $sk$ and message $m$ as input and outputs a signature $\sigma$.
    \item $b\gets Vefy(pk,m,\sigma)$ takes public key $pk$ and signature $\sigma$ as input and outputs a bit $b$, if $\sigma$ is a valid signature of message $m$, $b=1$, otherwise, $b=0$.
\end{itemize}

\subsubsection{Pseudo-random Generator (PRG)} In this article, we utilize a $PRG$ to facilitate the implementation of our algorithm. The $PRG$, $\mathbf{rv}\gets PRG(2^k, num, seed)$, that takes the security parameter $k$, the number of random numbers $num$, and the seed $seed$ as input, and outputs an array or vector, $\mathbf{rv}$, containing $num$ random numbers.

\section{Scheme} \label{section:our_scheme}

In this section, we first introduce two major algorithms, and then show the complete scheme.

\subsection{The Aggregation Algorithm Resistant to Malicious Pariticpant}\label{module:aggregation}

As attack methods continue to evolve, designing robust aggregation algorithms that can resist malicious participants is crucial for FL. In this paper, we adopt an aggregation algorithm to achieve robustness by detecting bad gradients with wrong directions uploaded by participants. The detailed flow of the algorithm is shown in Algorithm \ref{alg:aggregation}. We assume that $CS$ holds a benign small dataset and can calculate benign server gradients, and let $P_r$ be the set containing honest participants in round $r$ of training.

In the $r$-th round of joint training, $CS$ performs local training to generate server gradient $\mathbf{g}^S_r$.
After receiving local gradients uploaded by participants in $P$, $CS$ locally executes the aggregation algorithm.
As shown in steps 2 to 9 of the algorithm, if the cosine similarity between the local gradient $\mathbf{g}^i_r$ and the server gradient $\mathbf{g}^S_r$ is not positive, $\mathbf{g}^i_r$ is a bad gradient with wrong direction; otherwise, $\mathbf{g}^i_r$ is a benign gradient and $i$ is a honest participant, $CS$ will add $i$ to the set $P_r$. Note that $\mathbbm{1}(x)$ in the aggregation algorithm is a symbolic function, $\mathbbm{1}(x)=0$ if $x\le 0$, otherwise, $\mathbbm{1}(x)=1$. 

As shown in steps 11 to 14, at the end of this round, $CS$ aggregates benign local gradients to get the aggregated gradient $\mathbf{ag}_r$ that will be used to update the global model. Considering that participants may maliciously amplify gradients to mislead training, we can remove local gradients whose inner product with server gradient is too large.

\begin{algorithm}[hb]
\caption{Robust Aggregation.}
\label{alg:aggregation}
\begin{algorithmic}[1]
\REQUIRE In the $r$-round, the server gradient $\mathbf{g}^S_r$ of $CS$ and local gradient $\mathbf{g}^i_r$ of participant $i\in P$.\\
\ENSURE The aggregated gradient $\mathbf{ag}_r$.
    \STATE $\mathbf{ag}_r\gets \mathbf{g}^S_r$.
    \STATE $//$ \textit{Detect bad gradients.}
    \FOR{each $i\in P$}
        \STATE $//$ \textit{Calculate the inner product after normalizing the gradient.}
        \STATE $\mathbf{\Tilde{g}}^S_r\gets \frac{\mathbf{g}^S_r}{\Vert \mathbf{g}^S_r \Vert}, \mathbf{\Tilde{g}}^i_r\gets \frac{\mathbf{g}^i_r}{\Vert \mathbf{g}^i_r \Vert}$
        \STATE $d_i\gets \mathbbm{1}(\langle\mathbf{\Tilde{g}}^S_r,\mathbf{\Tilde{g}}^i_r\rangle).$
        \IF{$d_i == 1$}
        \STATE $//$ \textit{Participant $i$ provids benign local gradient.}
            \STATE $P_r.insert(p_i).$
        \ENDIF
        \STATE $//$ \textit{Aggregate benign gradients.}
        \STATE $\mathbf{ag}_r\gets \mathbf{ag}_r+d_i\cdot \mathbf{g}^i_r$.
    \ENDFOR
    \STATE $\mathbf{ag}_r\gets \frac{\mathbf{ag}_r}{|P_r|+1}$.
\RETURN $\mathbf{ag}_r$.
\end{algorithmic}
\end{algorithm}

\subsection{The Multiplication-related Pseudo-Random Vector Generation Algorithm}

In order to mask the gradients while allowing anyone to accurately verify that the aggregation algorithm is correctly performed in the $r$-th training round, the idea is to generate masked local gradients $\mathbf{mg}^i_{r,1}, \mathbf{mg}^j_{r,2}$ and masked server gradients $\mathbf{mg}^S_{r,i,1}, \mathbf{mg}^S_{r,2}$. As shown in Equation \ref{verifycos}, anyone can use $\mathbf{mg}^i_{r,1}$ and $\mathbf{mg}^S_{r,i,1}$ to accurately verify whether the detection of bad gradients is correct, and as shown in Equation \ref{verifyagg}, use $\mathbf{mg}^i_{r,2}$ and $\mathbf{mg}^S_{r,2}$ to verify whether the gradients of honest participants in round $r$ are correctly aggregated.
\begin{center}
\begin{equation}
    \begin{aligned}\label{verifycos}
        \Tilde{d}_i=\mathbbm{1}(\langle\mathbf{mg}^i_{r,1}, \mathbf{mg}^S_{r,i,1}\rangle)=d_i.
    \end{aligned}
\end{equation}
\end{center}
\begin{center}
\begin{equation}
    \begin{aligned}\label{verifyagg}
        \Tilde{\mathbf{ag}}_r=\frac{\mathbf{mg}^S_{r,2}+\sum_{j\in P_r}\mathbf{mg}^j_{r,2}}{|P_r|+1}=\mathbf{ag}_r.
    \end{aligned}
\end{equation}
\end{center}

In order to generate such two gradients, $\mathbf{mg}^i_{r,1}, \mathbf{mg}^S_{r,i,1}$, we consider that an algorithm can be implemented using a pseudo-random number generator that generates a pair of random vectors $\mathbf{rv}^i_r, \mathbf{rv}^S_r$ (3) of the same length as the gradient and with multiplicative correlation to be used as a mask.
\begin{center}
\begin{equation}
    \begin{aligned}\label{mul-related-rv}
        \mathbf{rv}^i_r\odot \mathbf{rv}^S_r=l\cdot\mathbf{1}.
    \end{aligned}
\end{equation}
\end{center}
where $\odot$ represents Hadamard product, which is the element-by-element multiplication of vectors, and $\mathbf{1}$ represents a vector whose elements are all $1$, so the multiplication of the elements at the corresponding positions in $\mathbf{rv}^i_r, \mathbf{rv}^S_r$ is equal to a positive number $l$, that is, $\mathbf{rv}^i_r[k]\cdot\mathbf{rv}^S_r[k]=l$.

Therefore, we set $\mathbf{mg}^i_{r,1}=\mathbf{rv}^i_r\odot \mathbf{g}^i_r, \mathbf{mg}^S_{r,i,1}=\mathbf{rv}^S_r\odot \mathbf{g}^S_{r,i}$ and satisfy the equation: $\mathbbm{1}(\frac{\langle\mathbf{mg}^i_{r,1}, \mathbf{mg}^S_{r,i,1}\rangle}{\Vert\mathbf{mg}^i_{r,1}\Vert\cdot \Vert\mathbf{mg}^S_{r,i,1}\Vert})=\mathbbm{1}(l\cdot\langle\mathbf{g}^i_r, \mathbf{g}^S_r\rangle)=\mathbbm{1}(\langle\Tilde{\mathbf{g}}^i_r, \Tilde{\mathbf{g}}^S_r\rangle)=\mathbbm{1}(\frac{\langle\mathbf{g}^i_r, \mathbf{g}^S_r\rangle}{\Vert\mathbf{g}^i_r\Vert\cdot \Vert\mathbf{g}^S_r\Vert})=d_i$.

However, when designing such a algorithm, we face the following problems. If $l$ is first randomly selected, then $\mathbf{rv}^i_r$ is randomly generated, and finally $\mathbf{rv}^S_r$ is calculated. In order to make $\mathbf{rv}^i_r$ and $\mathbf{rv}^S_r$ satisfy Equation 3, let each element $\mathbf{rv}^S_r[k]=\frac{l}{\mathbf{rv}^i_r[k]}$. However, $\mathbf{rv}^S_r[k]$ calculated in this way is likely to be an infinite decimal and will be truncated, which causes Equation \ref{verifycos} to fail to hold. Although the errors caused by the truncation of each element are small, after the massive multiplication and addition calculations in Equation \ref{verifycos}, these small errors will be aggregated and amplified, causing verification failure. If we think about it differently, we can generate $\mathbf{rv}^i_r$ first, then find a value $l$ that divides all the elements of the vector $\mathbf{rv}^i_r$, and use $l$ to compute another vector $\mathbf{rv}^S_r$. However, we find that this leads to even more serious upward overflow, i.e., $l$ is too big.

\begin{algorithm}[hb]
\caption{Pseudo-Random Vector Generation (PRVG).}
\label{alg:randomvector-gen}
\begin{algorithmic}[1]
\REQUIRE The random seed $seed$, the random number $\mathbf{rn}$ and $num$.\\
\ENSURE A vector $\mathbf{rv}$ with $num$ elements.
    \IF{$num\le 0$}
        \STATE BREAK.
    \ENDIF
    \FOR{$i\in[0,num)$}
        \STATE $sign\gets RandomSample(seed,i,\{1,-1\})$. $//$ \textit{The sign ($1$ or $-1$) of the $i$th element of the vector is sampled uniformly.}
        \STATE $\mathbf{rv}[i]\gets sign\cdot rn.$
    \ENDFOR
\RETURN $\mathbf{rv}$.
\end{algorithmic}
\end{algorithm}

Our solution idea is to first generate two random numbers $rn^i$, $rn^S$ with the same sign using PRG with input security parameter $k$, and make the positive number $l=rn^i\cdot rn^S$. Also, since it is just a multiplication of two random numbers, $l$ will not be very large and overflow. The random numbers and random seeds are then seeded into the pseudo-random vector generation (PRVG) algorithm \ref{alg:randomvector-gen} we designed to generate two random vectors $\mathbf{rv}^i_r$ and $\mathbf{rv}^S_r$ for the $r$th round of training. The key point here is that the algorithm $RandomSample(seed,i,\{1,-1\})$ used in PRVG is based on the PRG implementation and outputs the same sign as long as $seed$ and $i$ are the same. Thus multiplication of elements at the same position in $\mathbf{rv}^i_r$ and $\mathbf{rv}^S_r$ results in $l$, which means that the two vectors satisfy Equation 3.

Secondly, we use the idea of zero sharing to generate $\mathbf{mg}^j_{r,2}, \mathbf{mg}^S_{r,2}$. First, divide the all-zero vector $\mathbf{0}$ into $|P_r|+1$ shares {$\mathbf{zv}^S, \mathbf{zv}^j,\ j\in P_r$}. Then, $\mathbf{mg}^j_r$ and $\mathbf{mg}^S_r$ can be regarded as additive shares of $\mathbf{ag}_r$. Therefore, according to the homomorphism of additive secret sharing, we can get $\frac{(\mathbf{zv}^S+\mathbf{mg}^S_r)+\sum_{j\in P_r}(\mathbf{zv}^j+\mathbf{mg}^j_r)}{|P_r|+1}=\frac{\mathbf{g}^S_r+\sum_{j\in P_r}\mathbf{g}^j_r}{|P_r|+1}=\mathbf{ag}_r$. Let $\mathbf{mg}^j_{r,2}=\mathbf{zv}^j+\mathbf{mg}^j_r, \mathbf{mg}^S_{r,2}=\mathbf{zv}^S+\mathbf{mg}^S_r$, then Equation \ref{verifyagg} is established.

\subsection{The Publicly Auditable and Privacy-Preserving Federated Learning Scheme} \label{schemedetails}

In this section, we show the details of the scheme in Figure \ref{scheme: FLscheme}. Based on the algorithm introduced above, in our federated learning scheme, $CS$ and participants generate random vectors for masking private information. Then during the model training phase, $CS$ computes the masked gradient with participants. After training, $CS$ packages these gradients and masked models into training record $TRCD$ and publishes it on the blockchain, allowing anyone to audit the correctness of the training process without leaking privacy.

$CS$ then deploys model $w$ into actual services and does not expose $w$ to clients. In order to enhance trust in the model, clients can retrieve $w$'s training record $TRCD$ on the blockchain and trace the training process. As shown in Equation \ref{verifycos} and \ref{verifyagg}, clients can check whether the process of $CS$ detecting malicious actors, aggregating gradients from honest participants, and updating the model is correct without knowing the gradients. The audit process is as follows:
\begin{enumerate}
    \item Clients can obtain the set $P_r$ of participants who uploaded benign gradients in round $r$ according to Equation \ref{verifycos} and, according to Equation \ref{verifyagg}, calculate the aggregated gradient masked by $\mathbf{rv}^S_{w,r}$, $\mathbf{mag}_r=\frac{\mathbf{mg}^S_{r,2}+\sum_{j\in P_r}\mathbf{mg}^i_{r,2}}{|P_r|+1}$.
    \item Then, clients can recover the masked initial global model $\mathbf{w}_0=-\mathbf{rv}^S_w+\mathbf{w}+\eta\cdot\sum^R_{r=1}\mathbf{mag}_r$.
    \item Finally, clients verify that $\mathbf{w}_0$’s signature $\sigma^S_w$ is correct.
\end{enumerate}

Herein, it is evident that the client has completed tracing and verifying the training process of the global model but does not know the gradients and models involved in the training. Therefore, the client will trust that the global model deployed in the service is correctly trained, and that the privacy of $CS$ and participants has not been compromised.

\begin{figure*}
\centering
\fbox{\parbox[c][][t]{0.99\textwidth}{
\textbf{Parameters:} Security parameter $k$.\\
\textbf{Initialization:} Generate keys and pre-train the global model.
  \begin{itemize}
    \item[-] Participant $i$ generates a pair of keys, $(sk^i,pk^i)\gets GenKey(1^k)$ and $CS$ generates $(sk^S,pk^S)\gets GenKey(1^k)$.
    \item[-] At the same time, $CS$ defines the machine learning task $(model\ structure,\ number\ of\ training\ rounds:\ R,\ learning\ rate:\ \eta)$ and pre-trains the global model $\mathbf{w}_0$.
  \end{itemize}
\textbf{Preprocessing:} Distribute random numbers.
  \begin{itemize}
    \item[-] $CS$ generates $\mathbf{rv}^S_w,\ seed^i,\ \mathbf{zv}^i,\ rn^i,\ rn^S_i,\ l^i$, where $\mathbf{rv}^S_w$ is a random vector used to mask the model, $\mathbf{zv}^i$ is a vectorized zero share used to mask the local gradient of participant $i$, and $rn^i,\ rn^S_i$ is two random numbers with the same sign, and $l^i=rn^i\cdot rn^S_i$ is a positive number.
    \item[-] $CS$ divides $\mathbf{rv}^S_w$ into $R$ shares $\mathbf{rv}^S_{w,r}$, $\sum^R_{r=1}\mathbf{rv}^S_{w,r}=\mathbf{rv}^S_w$, and sends $seed^i,\ \mathbf{zv}^i,\ rn^i$ to participant $i\in P$ and publishes the training task and signature $\sigma^S_w$ of the initial global model $\mathbf{w}_0$ on the blockchain, $\sigma^S_w\gets Sign(sk^S, \mathbf{w}_0)$.
  \end{itemize}
\textbf{Model training:} $CS$ schedules participants to perform tasks and records the model training process.
  \begin{itemize}
    \item[-] In the $r$-th round of training, $0<r\le R$, $CS$ first releases the latest global model $w_{r-1}$ to participants. Then, $CS$ and participant $i$ perform local training on $\mathbf{w}_{r-1}$ to obtain server gradient $\mathbf{g}^S_r$ and local gradient $\mathbf{g}^i_r$ respectively.
    \item[-] Each participant $i$ calculates $\mathbf{rv}^i_r\gets PRVG(seed^i+r, rn^i, num)$, and $CS$ calculates $\mathbf{rv}^S_{r,i}\gets PRVG(seed^i+r, rn^S_i, num)$, $num$ is the number of elements of the local gradient and $PRVG$ as shown in Algorithm \ref{alg:randomvector-gen}.
    \item[-] Each participant $i$ computes masked local gradients $\mathbf{mg}^i_{r,1}=\mathbf{rv}^i_r\odot \mathbf{g}^i_r,\ \mathbf{mg}^i_{r,2}=\mathbf{zv}^i+\mathbf{g}^i_r$, and sends local gradient $\mathbf{g}^i_r$ and signatures $\sigma^i_{r,mg1}\gets Sign(sk^i, \mathbf{mg}^i_{r,1}),\ \sigma^i_{r,mg2}\gets Sign(sk^i, \mathbf{mg}^i_{r,2})$ to $CS$.
    \item[-] $CS$ computes $\mathbf{mg}^i_{r,1},\ \mathbf{mg}^i_{r,2}$ and verifies the correctness of $\sigma^i_{r,mg1},\ \sigma^i_{r,mg2}$, and executes the Robust Aggregation Algorithm \ref{alg:aggregation} to obtain $\mathbf{ag}_r$, where $\mathbf{ag}_r$ is aggregated gradient. $CS$ sets $P_r$ to the set of honest participants in round $r$. In this article, participant $i$ is considered a malicious participant if the signature verification fails or the cosine similarity between the local gradient $\mathbf{g}^i_r$ and the server gradient $\mathbf{g}^S_r$ is negative ($d_i=0$).
    \item[-] $CS$ updates the global model $\mathbf{w}_r\gets \mathbf{w}_{r-1}-\eta\cdot \mathbf{ag}_r$ and calculates  $\mathbf{zv}^S_r=-\sum_{j\in P_r}\mathbf{zv}^j_r$, $\mathbf{mg}^S_{r,i,1}$, $\mathbf{mg}^S_{r,2}$, where $\mathbf{zv}^S_r,\mathbf{zv}^j_r,j\in P_r$ are shares of the all-zero vector $\mathbf{0}$, masked server gradients $\mathbf{mg}^S_{r,i,1}=\mathbf{rv}^S_{r,i}\odot \mathbf{g}^S_r$, $i\in P$, $\mathbf{mg}^S_{r,2}=\frac{(|P_r|+1)\cdot\mathbf{rv}^S_{w,r}}{\eta}+\mathbf{zv}^S_r+\mathbf{g}^S_r$.
    \item[-] $CS$ then packs the masked gradients into $mlg_r,msg_r,trcd_r$, where $mlg_r=\{\mathbf{mg}^i_{r,1},\mathbf{mg}^j_{r,2},\sigma^i_{r,mg1},\sigma^i_{r,mg2}\}$ and $msg_r=\{\mathbf{mg}^S_{r,i,1}\}$, $,\ i\in P,j\in P_r$, are the sets of masked local gradients and server gradients respectively and $trcd_r=\{mlg_r,msg_r\}$.
    \item[-] If $r\le R$, continue model training, else, the training ends, $CS$ obtains the trained model $\mathbf{w}$ and computes masked model $\mathbf{mw}=-\mathbf{rv}^S_w+\mathbf{w}$. 
    \item[-] At last, $CS$ computes $rt^i$ about each $l^i$, and packs the training records of each round into a total training record $TRCD=(trcd_1,\cdots,trcd_R,\ \mathbf{mw},\ RT)$. $RT=(rt^1,rt^2,\cdots,rt^|P|)$ and for each $l^i$, $l^j$, $rt^i\cdot l^i=rt^j\cdot l^j$. At last $CS$ uploads $TRCD$ to the blockchain.
\end{itemize}
}}
    \caption{Publicly Auditable, Privacy-Preserving and Robust Federated Learning Based on Blockchain and Zero Sharing.}
    \label{scheme: FLscheme}
\end{figure*}
    
\subsection{Security Analysis}
As can be seen from the previous discussion, compared with existing research results on trusted federated learning, our solution simultaneously achieves public auditability, privacy protection and training robustness.
We discuss in this section how our scheme achieves privacy and public auditability under the security model. 

\subsubsection{Privacy} $\mathbf{mg}^i_{r,1}$, $\mathbf{mg}^S_{r,i,1}$, $\mathbf{mg}^j_{r,2}$, $\mathbf{mg}^S_{r,2}$, $\mathbf{mag}_r$ and $-\mathbf{rv}^S_w+\mathbf{w}$, $i\in P, j\in P_r, r=1,\cdots,R$ in the training record $TRCD$ will not reveal the private information of the local gradients, server gradients, aggregated gradients and global model to participants and clients.

$\mathbf{rv}^i_r$, $\mathbf{zv}^j$ and $\mathbf{rv}^S_{r,i}$, $\mathbf{zv}^S_r$, $\mathbf{rv}^S_{w,r}$ are random vectors generated by $CS$ using PRG and used to mask $\mathbf{g}^j_r$ of participant $j\in P_r$ and $\mathbf{g}^S_r$, $\mathbf{ag}_r$.
Therefore, based on the security of $PRG$, it is difficult for participant $i$ and clients to guess the multiplication-related random vectors $\mathbf{rv}^j_r$, $\mathbf{rv}^S_{r,j}$ of participant $j$ and $CS$, and since the elements in $\mathbf{rv}^j_r$ are not all equal, the magnitude and direction of participant $j$’s local gradient will not be leaked to other participants. The same is true for the gradient of the server, although the participant owns a portion of the multiplication-related random vector. Thus, $\mathbf{mg}^i_{r,1}$, $\mathbf{mg}^S_{r,i,1}$ do not leak any information about the gradient, even if it is an element of the gradient.

In the same way, $\mathbf{mg}^i_{r,2}$ and $\mathbf{mg}^S_{r,2}$, $\mathbf{mag}_r$ will not disclose information of gradients $\mathbf{g}^j_r$, $\mathbf{g}^S_r$ to other participants or clients.
Second, $\mathbf{rv}^S_w=\sum^R_{r=1}\mathbf{rv}^S_{w,r}$ is used to mask the final trained global model $\mathbf{w}$.
Based on the security of additive ($R,R$)-secret sharing, that is any $t$ ($t<R$) shares will not reveal any information about the secret $\mathbf{rv}^S_w$, $\mathbf{mw}$ will not disclose $\mathbf{w}$ to anyone.

\subsubsection{Public Auditability} The training record $TRCD$ generated in our solution support any third party to audit the model training, that is, to accurately trace the model training process and verify the correctness.

The audit process has been explained in Section \ref{schemedetails}, and we will analyze its correctness here. First, the detection of bad gradients can be verified by Equation \ref{verifycos} for correctness. The aggregated gradient can be audited as the following formula.
\begin{center}
    \begin{align*}            
        \mathbf{mag}_r&=\frac{\frac{(|P_r|+1)\cdot\mathbf{rv}^S_{w,r}}{\eta}+\mathbf{zv}^S_r+\mathbf{g}^S_r+\sum_{j\in P_r}\mathbf{zv}^j+\mathbf{g}^j_r}{|P_r|+1}\\
        \nonumber
        &=\frac{\mathbf{rv}^S_{w,r}}{\eta}+\frac{(\mathbf{zv}^S_r+\sum\limits_{j\in P_r}\mathbf{zv}^j)+(\mathbf{g}^S_r+\sum\limits_{j\in P_r}\mathbf{g}^j_r)}{|P_r|+1}\\
        \nonumber
        &=\frac{\mathbf{rv}^S_{w,r}}{\eta}+\mathbf{ag}_r.\\
        \nonumber
        \mathbf{w}_0=&-\mathbf{rv}^S_w+\mathbf{w}+\eta\cdot(\sum^R_{r=1}\frac{\mathbf{rv}^S_{w,r}}{\eta}+\mathbf{ag}_r)\\
        \nonumber
        =&-\mathbf{rv}^S_w+\mathbf{w}+\mathbf{rv}^S_w+\eta\cdot\sum^R_{r=1}\mathbf{ag}_r=\mathbf{w}+\eta\cdot\sum^R_{r=1}\mathbf{ag}_r.
        \nonumber
    \end{align*}
\end{center}
Then, as shown in the above formula, anyone can trace the model update process to obtain the initial model $\mathbf{w}_0$.

Finally, verify whether the retroactively restored model $\mathbf{w}_0$ is the initial model signed by $CS$ before the start of training to determine whether the gradient aggregation and model update are correct.

Furthermore, it is worth mentioning that for verifying whether $CS$ clipped the gradient suspected to be maliciously amplified. The third party can compute $\Tilde{inp}^i=\langle\mathbf{mg}^i_{r,1}, \mathbf{mg}^S_{r,i,1}\rangle$ and then use $rt^i$ in $TRCD$ to deflate $\Tilde{inp}^i$. That is, for any $\Tilde{inp}^i$, $\Tilde{inp}^j$, $rt^i\cdot \Tilde{inp}^i=rt^i\cdot l^i\cdot \langle\mathbf{g}^i_{r,1}, \mathbf{g}^S_{r,i,1}\rangle$, $rt^j\cdot \Tilde{inp}^j=rt^j\cdot l^j\cdot \langle\mathbf{g}^j_{r,1}, \mathbf{g}^S_{r,j,1}\rangle$ and $rt^i\cdot l^i=rt^j\cdot l^j$ holds. A sort can then be performed to check which local gradient has the largest inner product with the server gradient.

\section{Implementation and Evaluation} \label{section:evaluation}

In this section, we evaluate the performance of our scheme. We implemented our solution based on Python 3.8.15 and tensorflow 2.11.0 on a server running Ubuntu 22.04 equipped with Intel Core CPU i7-12700F 2.10 GHz and NVIDIA RTX A6000 GPU. The protocol specifies that $CS$ and 20 FL participants collaborate to train a classification model on the MNIST dataset \cite{lecun1998mnist}. The dataset will be randomly divided into 21 subsets. We trained the model with different methods in the presence of 0 to 9 malicious participants who randomly replaced the labels of the training data. The training parameters are as follows: the security parameter $k$ is 48, the training rounds $R$ is 500, the learning rate is 0.5, and the mini-batch size is 64.

\begin{figure}[htbp]
\centering
\includegraphics[width=\linewidth]{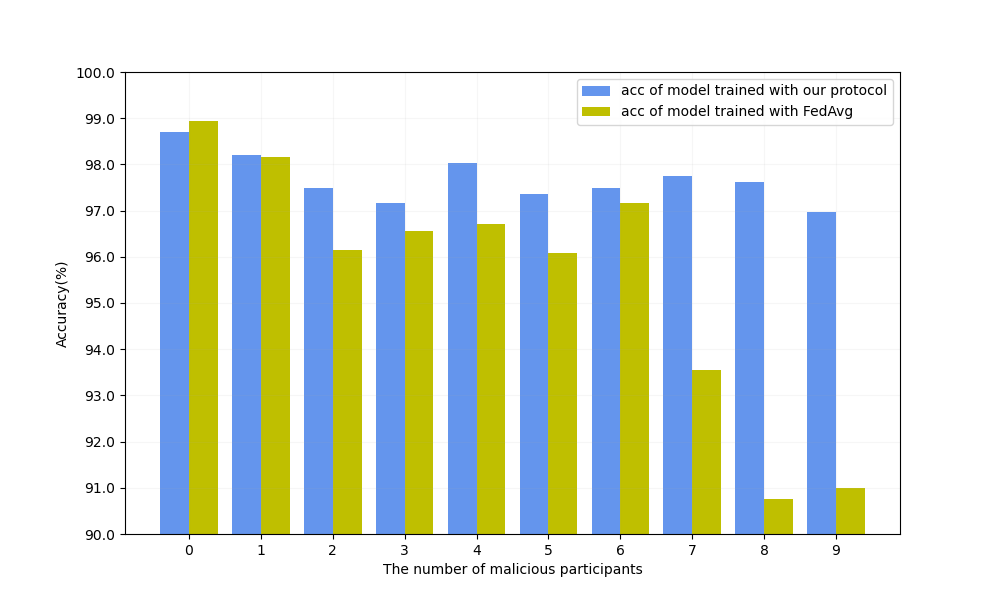}
\caption{The model accuracy of schemes under different numbers of malicious participants.}
\label{Eva1}
\end{figure}

\begin{figure}[htbp]
\centering
\includegraphics[width=\linewidth]{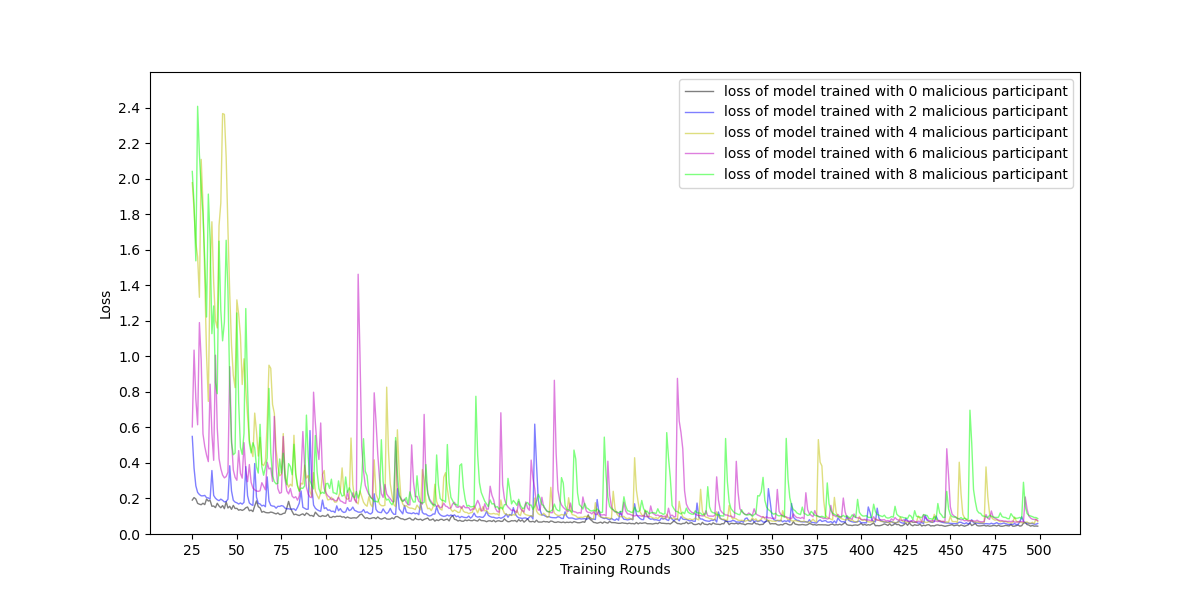}
\caption{The training loss of schemes under different numbers of malicious participants.}
\label{Eva2}
\end{figure}

As shown in Figure \ref{Eva1}, we can see that as the number of malicious participants increases, the accuracy of the model trained by our scheme decreases slowly, but the accuracy of the model trained by the less robust scheme, FedAvg \cite{FL:FedAvg/McMahanM17}, decreases faster. The accuracy of the model trained by our scheme is higher than that of the model trained by FedAvg, as shown in Table \ref{Eva1_num} for details, where "\textbf{0 MP}" represents the case of 0 malicious participants.
In Figure \ref{Eva2}, we compare the convergence speed of model loss values in different cases. It is obvious that as the number of malicious participants increases, the loss value fluctuates more and converges more slowly.

\begin{table}[htbp]
\centering
\caption{Accuracy of the model of our scheme and FedAvg}
    \setlength{\tabcolsep}{1.6mm}{
        \resizebox{\linewidth}{!}{
        \begin{tabular}{|c|c|c|c|c|c|c|c|c|c|c|}
        \hline
         & \textbf{0 MP} & \textbf{1 MP} & \textbf{2 MP} & \textbf{3 MP} & \textbf{4 MP} & \textbf{5 MP} & \textbf{6 MP} & \textbf{7 MP} & \textbf{8 MP} & \textbf{9 MP} \\
        \hline
        Our & 98.70\% & 98.21\% & 97.49\% & 97.16\% & 98.04\% & 97.35\% & 97.49\% & 97.74\% & 97.61\% & 96.97\%\\
        \hline
        FedAvg & 98.94\% & 98.16\% & 96.15\% & 96.56\% & 96.71\% & 96.09\% & 97.16\% & 93.54\% & 90.76\% & 90.99\%\\
        \hline
    \end{tabular}}
    
    \label{Eva1_num}}
\end{table}

In terms of total time overhead, without considering the network delay, the time overhead includes the time cost for each participant to generate two signatures and compute $\mathbf{mg}^i_{r,1}$, $\mathbf{mg}^i_{r,2}$, and $CS$ to verify each participant's signature in each round of training, which are 0.2113s and 0.0231s respectively. At the same time, we measured that the time overhead of executing the $PRVG$ algorithm is 0.1692s. The time overhead of the generation of $zv^i$ is 0.0562s, but this can be completed before the start of training. In addition, before training, the communication overhead caused by $CS$ sending random numbers etc. to participant $i$ is 0.554 MB, and in each training round, the communication overhead of participants is 1.1196 MB.

\section{Conclusion}\label{section:conclusion}
In this paper, we present a publicly auditable and privacy-preserving federated learning scheme that outputs high-quality and trustworthy models that can be deployed to real-world applications, even if malicious participants participate in training. 
Specifically, we present a robust aggregation algorithm to ensure that gradients with wrong directions uploaded by malicious participants do not affect the accuracy of the model. In addition, we designed an $PRVG$ algorithm and combined it with zero sharing and blockchain to generate training records that support auditing of the training process by anyone. Experiments show that the scheme is efficient. In future work, we intend to further improve the training accuracy.

\section*{Acknowledgment}
This work was supported by the Key-Area Research and Development Program of Guangdong Province(Grant Nos. 2020B0101090004), the National Natural Science Foundation of China (62072215, 72374058, U22B2028,U23A20303), Guangdong-Hong Kong Joint Laboratory for Data Security and Privacy Preserving (Grant No. 2023B1212120007),  the Guangdong Key Laboratory of Data Security and Privacy Preserving (Grant No. 2023B1212060036).

\ifCLASSOPTIONcaptionsoff
  \newpage
\fi

\bibliographystyle{IEEEtran}
\bibliography{reference}

\end{document}